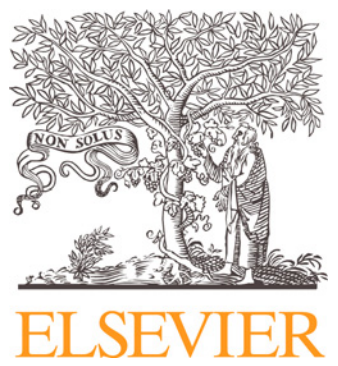

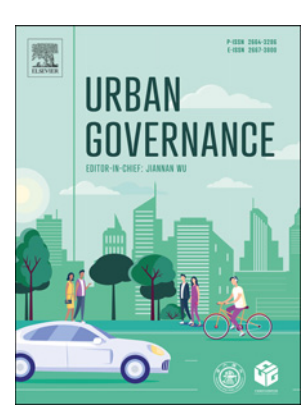

# AI and the Transformation of Accountability and Discretion in Urban Governance

Stephen Goldsmith[a], Juncheng Yang[a,b,*]

[a] Bloomberg Center for Cities, Harvard University, 79 JFK Street, Cambridge, MA 02138, United States
[b] Department of Urban Planning and Design, Harvard University Graduate School of Design, Cambridge, MA 02138, USA

ABSTRACT

This paper offers a conceptual analysis of the transformative role of artificial intelligence (AI) in urban governance, focusing on how AI can reshape the relationship between bureaucratic discretion and accountability. Drawing on public administration theory and algorithmic governance research, the study argues that AI does not simply restrict or enhance discretion but redistributes it across institutional levels, professional roles, and citizen interactions. While primarily conceptual, this paper uses illustrative cases to show that AI can strengthen managerial oversight, improve service delivery consistency, and expand citizen access to information. These changes affect different forms of accountability: political, professional, and participatory, while introducing new risks, such as data bias, algorithmic opacity, and fragmented responsibility across actors. In response, the paper introduces the concept of *accountable discretion* and proposes guiding principles, each linked to actionable measures: equal AI access, adaptive administrative structures, robust data governance, proactive human-led decision-making, and citizen-engaged oversight. This study contributes to the AI governance literature by moving beyond narrow concerns with perceived discretion at the street level, highlighting instead how AI transforms rule-based discretion across governance systems. It also reframes the trade-off between discretion and accountability as a dynamic and evolving relationship shaped by algorithmic systems and institutional practices.

## 1. Introduction

### 1.1. The increasing role of data-driven approaches in urban governance

Integrating digital tools such as ubiquitous personal devices, the Global Positioning System (GPS), and various Internet of Things (IoT) devices into the urban environment is reshaping citizens' experiences and enabling new forms of data collection and analysis for urban governance. Digital systems facilitate the convergence of data generated from various sources, including networked IoT sensors, cross-agency transactions, individuals and their communities. These systems enhance government performance and the quality of urban life by enabling city-level data integration, agency-tailored interfaces and systems for management, and interactive digital twinning interventions (Appio et al., 2019; Hawken et al., 2020; Kitchin et al., 2015; O'Reilly, 2011; Rabari & Storper, 2015; Schrotter & Hürzeler, 2020; Yang & Rong, 2024). For example, Singapore's Smart Nation Sensor Platform harnesses real-time data from thousands of interconnected sensors to optimize public transportation, monitor environmental quality, and enhance public safety (MDDI, 2024). Similarly, Barcelona has deployed extensive IoT networks that track air quality and traffic flow, empowering municipal authorities to make agile, informed decisions that improve residents' daily lives (Barcelona City Council, 2025). In Amsterdam, the use of AI-powered analytics within smart grid systems has assisted the municipality delivering more efficient solutions for energy management and waste reduction (Gemeente Amsterdam, 2024). Cities like Boston, Portland, and San José are piloting digital curb management platforms and IoT-enabled infrastructure to better regulate street space and improve pedestrian safety (Goldsmith, 2025).

Technological advancements prompt paradoxical impacts on bureaucracy. On the one hand, aligning with Weber's view that technology as an enabler for bureaucratization (1946), emerging data- and AI-driven innovations might reinforce traditional bureaucratic structures and practices (Meijer et al., 2021; Newman et al., 2022). On the other hand, urban governance faces complex, "wicked" problems (Rittel & Webber, 1973), necessitating new governance models that utilize flexible and cross-agency institutional arrangements (Stead & Meijers, 2009). Digital technology has facilitated this shift towards networked organizational structures (Greenwood & Lawrence, 2005), creating odds with the Weberian bureaucratic model which traditionally emphasizes rigid rules and standardized processes. In particular, this digital transformation enhances agility and flexibility for public employees, enabling

* Corresponding author.
*E-mail addresses:* stephen_goldsmith@hks.harvard.edu (S. Goldsmith), juncheng_yang@gsd.harvard.edu (J. Yang).

them to address societal needs more effectively (Goldsmith & Gardner, 2022).

### 1.2. Rethinking bureaucratic discretion

Building upon the vast amount of data, the rapidly growing use of AI promises a further transformation of urban governance. The intertwined relationship between humans and AI in bureaucracy resonates with the concept of *accountable discretion*, suggesting that data-driven decision-making can enhance public outcomes without compromising democratic controls (Goldsmith & Crawford, 2014). In the context of AI, deploying AI in bureaucratic systems may further create opportunities to augment the discretion of bureaucrats while upholding stringent accountability standards due to AI's enabling effects on the data-processing capacities of government employees.

While much of the existing literature has focused on the perceived or "felt" discretion among street-level bureaucrats (Busch et al., 2018; Flügge et al., 2021; Brayne & Christin, 2021; de Boer & Raaphorst, 2023; Giest & Klievink, 2024). In contrast, this paper distinguishes itself by exploring the formal, rule-based dimensions of discretion embedded within governance structures and processes, especially when facing the emergence of highly interactive AI applications following the rise of Generative AI (GenAI). By integrating perspectives into the institutional mechanisms that define discretion and accountability, this paper provides a more comprehensive understanding of how AI applications could reshape decision-making processes. Moreover, considering variations across frontline staff and senior management, this paper aims to reveal how different actors adapt and reallocate their discretionary powers in response to AI-enabled transformations.

However, as AI broadens discretionary capacity, it must be managed carefully to ensure that enhanced flexibility does not compromise ethical standards or democratic principles. This change calls for a further rethinking of traditional accountability frameworks to accommodate the new human-AI interactions, hence ensuring that discretion is exercised in a manner that is both responsible and aligned with ethical values.

To further elaborate on these themes, the remainder of this paper is organized as follows: Section 2 reviews the traditional trade-off between discretion and accountability in governance, establishing the theoretical framework for our analysis. Section 3 investigates how AI impacts discretion and accountability in urban governance, focusing on AI-assisted approaches and their implications for decision-making and oversight. Section 4 outlines a set of guiding principles for AI-enhanced governance in urban policy. Rather than offering a rigid framework, this section presents key strategies, including equitable AI deployment, adaptive administrative frameworks, robust data governance, transparent human–AI collaboration, and active citizen engagement, to help city governments harness AI's transformative potential while ensuring accountability, transparency, and ethical oversight. Section 5 concludes the paper and outlines directions for future research.

## 2. Literature review: discretion, accountability, and the trade-off

*Discretion* encompasses the authority to make decisions in specific situations within a bureaucrat's authority. The early definition of discretion was broad and did not attempt to categorize particular types of discretion according to tasks or contexts, considering discretion as a public officer's freedom "… to choose among possible courses of action and inaction" within its effective authority limits (Davis, 1969, p.4). Michael Lipsky's seminal work, *Street-level Bureaucracy: Dilemmas of the Individual in Public Services* (1980), articulates discretion specifically for frontline bureaucrats, emphasizing that their discretionary power is an essential component of implementing policy and shaping service delivery, especially in situations where rules may be ambiguous, complex, or contradictory. Discretion allows these bureaucrats to interpret and apply to fit the specific contexts they encounter daily with citizens (Hupe & Hill, 2007).

Building on Lipsky's studies, subsequent literature has recognized the significant role of street-level bureaucrats' discretionary actions in policy implementation. These bureaucrats are acknowledged as professionals who use their judgment to interpret and apply policies in ways that directly impact public service delivery (Hupe, 2013; Thomann et al., 2018). Their choices are shaped by their understanding of policy goals and political agenda (May & Winter, 2009), professional knowledge (Evans, 2011), as well as their own understanding of the relationships among citizen clients, coworkers, and the system (Maynard-Moody & Musheno, 2000). Discretion can also affect how meaningful frontline workers perceive their decisions to be (even if citizens might feel differently) (Tummers & Bekkers, 2014; Thomann et al., 2018; Wang et al., 2022). By contrast, more rigid policies that leave little room for discretion, such as mandatory arrest policies, would usually lead to more predictable actions, even when those compulsory actions generate harm or unfortunate outcomes. Because many frontline situations are non-programmed, the ability of bureaucrats to make discretionary decisions is essential for maintaining the flexibility and responsiveness required in complex urban governance (Hupe & Hill, 2007; Hupe, 2013).

Concerns about accountability accompany the discussion of discretion. The traditional definition of accountability highlights the concept's relational aspects. It emphasizes an agent's obligations to explain and justify their actions to another entity, which could be a person, an organization, or the public in general, holding the power to question and evaluate the agent's actions (Bovens, 1998, 2007, 2010; Mulgan, 2000; Mulgan, 2003). Traditionally, organizations ensure accountability through institutional mechanisms such as administrative procedures, which formalize processes and standards to be followed by bureaucrats (McCubbins et al., 1987) as well as democratic oversight, which allows citizens to be engaged in monitoring and evaluating government actions (McCubbins & Schwartz, 1984).

Accountability and discretion are often described to be in tension. High levels of discretion challenge the liberal framework of separation of powers of law-making and enforcement (Pires, 2011), leading to a perception of inadequate accountability. A key element underlying this trade-off is standardization. As discretion increases, decision-making processes tend to become less uniform. When decisions are sometimes tailored to specific contexts, the variability introduced makes it more challenging for oversight bodies to apply uniform accountability measures, such as audits or performance reviews, which rely on predictable, standardized processes.

Conversely, excessive control or oversight can undermine street-level bureaucrats' discretion to perform their duties effectively. That can reduce the efficiency of government operations and the quality of services (Hupe & Hill, 2007). Furthermore, overly rigid bureaucratic procedures can erode public trust by preventing sensible decisions or creating cumbersome or confusing administrative processes, which hinder governmental agencies' performance (Handler, 1986; Bryner, 1987; as cited in Pires, 2011). Research has also shown that overly restricted discretion might lower frontline bureaucrats' willingness to implement policies (Tummers & Bekkers, 2014; Thomann et al., 2018; Wang et al., 2022).

However, it is essential to note that this tension does not imply that accountability and discretion are fundamentally incompatible. Street-level bureaucrats remain legally and normatively bound to justify their actions, but higher levels of discretion expand the range of possible decisions, increasing variability and thereby complicating standardized forms of monitoring and evaluation. Much of this foundational work, however, predates the rise of algorithmic service delivery and decision support. As AI systems increasingly permeate public administration (Engstrom & Ho, 2021; Straub et al., 2023), the assumptions underpinning this discretion-accountability trade-off merit reconsideration. These tools may not simply constrain human choice but could redistribute it by shifting interpretive authority and monitoring responsibility across organizational levels, between humans and machines, and between public authorities and citizens. This shift calls for expanded dis-cussions of how discretion and accountability should operate in such hybrid decision-making environments. This paper addresses these theoretical gaps by re-examining the discretion–accountability relationship as a dynamic process rather than a zero-sum exchange, under conditions of algorithmic governance, setting the stage for subsequent discussions on achieving a more adaptive and transparent governance model.

## 3. AI's impact on bureaucratic discretion and accountability: a conceptual exploration

### 3.1. AI for urban governance

In urban governance, the rapid development of AI technology may lead to a bundle of transformative tools that city managers and frontline staff adopt to address context-specific challenges. AI encompasses a variety of technologies and techniques, including machine learning, neural networks, and natural language processing to solve "complex goals" that typically involve tasks similar to human intelligence, such as learning, reasoning, problem-solving, perception, and language understanding (Russell & Norvig, 2021; Tegmark, 2017). The following sections examine how AI's practical applications drive enhanced operational efficiency and a fundamental transformation in balancing bureaucratic discretion and accountability in urban governance.

Urban governments have been increasingly incorporating AI to enhance operational efficiency and decision-making. OECD (2024) reports that 67 % of OECD countries already use AI to improve service delivery, predominantly through tools such as chatbots, automated document-processing systems, and predictive analytics for resource management, while only 30 % apply AI for policy-making. Notable examples include municipal chatbots for citizen inquiries or internal workflows (Digital Dubai, 2023; Seoul Metropolitan Government, 2021; SNDGO, 2023) and the use of AI as a data analyst in the U.S. state and local agencies (Edinger, 2024; Rueter, 2024).

GenAI represents one of the latest innovations being explored by urban governments. Although GenAI is often discussed in terms of its technical capabilities, such as generating text, images, or predictive insights (Ouyang et al., 2022; Cao et al., 2023), the transformative feature is its interactive capacity. It allows a broader range of stakeholders, including citizens and non-technical public officials, to tackle tasks ranging from document summarization to data analytics and even coding through conversational languages. This change may significantly expand AI's applicability in urban governance processes.

Beyond expanding access, GenAI allows users to prompt, refine, and interpret responses iteratively. This dialogic mode of interaction does not replace human judgment; instead, it may enhance discretionary decision-making by making it more reflexive and context-sensitive. Empirical findings provide a glimpse of this shift. Boulus-Rødje et al. (2024), for instance, illustrate how users develop iterative working relationships with the GenAI; these working relationships enhance users' understanding of both the AI's limitations and their assumptions. Similarly, Law and Varanasi (2025) show how workers in diverse sectors enact a new managerial responsibility to verify or override GenAI outputs. Both examples suggest that GenAI may support more informed and traceable decisions that build upon users' evolving judgment and GenAI's refined responses.

As urban governance integrates more AI applications, these dynamics open up new challenges and opportunities for aligning tech-driven discretionary practices with established public values (see Moore, 2013). While AI applications could increase efficiency and scale, their true significance may lie in enabling officials across government levels to exercise more discretion in more informed, traceable, and reflexive ways. This shift may advance positive outcomes without losing accountability. These considerations lay the groundwork for the analysis in Section 3 and the principles discussed in Section 4, which aim to balance the transformative potential of AI with the essential needs for ethical oversight and accountable discretion in urban governance.

### 3.2. Understanding data-driven and AI-assisted approaches' impacts on discretion

The discussion on AI's impact on discretion draws parallel with the "curtailment-vs-enablement" debate regarding Information and Communication Technology's (ICT) effects on bureaucracy and discretion. On one side of the debate, scholars like Snellen (1998), Zuurmond (1998), Barth and Arnold (1999), and Bovens and Zouridis (2002) argued that the expanded use of early ICT applications may curtail discretion by automating tasks and reducing the need for human judgment. This reduction could decrease in discretionary power for human agents, potentially replacing street-level bureaucrats with data-driven algorithms.

In contrast, the "enablement" perspective posits that ICT tools enhance the ability of street-level bureaucrats by improving data-processing capacities and facilitating more informed decisions. This view suggests that the emergence of ICT tools may empower discretionary decision-making rather than suppressing human discretion. When bureaucrats embrace digital tools in daily operations, they still deem discretionary power an indispensable component of governance (Busch et al., 2018; Busch & Eikebrokk, 2019). In addition, the complexity of tasks in public administration also makes it difficult for digital systems to capture the full scope of frontline social works (Buffat, 2015; Nagtegaal, 2021). For instance, Jorna and Wagenaar's (2007) empirical study on ICT and subsidy allocation in the Netherlands demonstrated that while the ICT-driven monitoring records social workers' decisions, it also omits information about how much discretion a social worker exercised. This study further reinforces the point that discretion remains an impactful component in governance despite the broad deployment of digital technologies. Overall, literature has provided evidence for both enablement and curtailment theories when addressing ICT's effects on discretion, reaching a consensus that ICT's impacts on discretion are highly dependent on the nature and contexts of the tasks (Buffat, 2015; Busch & Henriksen, 2018).

When addressing the emergence of AI technology, literature recognizes a similar co-existence of curtailment and enablement effects, with specific task characteristics, such as the level of discretion required, complexity, or repetitiveness, playing a critical role (Gillingham et al., 2025). However, AI's effect on tasks is likely to vary. It reduces discretion in routine activities by automating processes, yet it can enhance decision-making in complex cases by offering high-quality data and predictive insights (Young et al., 2019; Bullock, 2019; Bullock et al., 2020).

#### 3.2.1. Repositioning discretion rather than replacing discretion

That being said, as AI presents the opportunity to reallocate routine tasks, it allows bureaucrats to concentrate on complex decisions that require human judgment. This repositioning reflects the hybrid nature of governance where AI supports human decision-making without entirely removing human judgment.

The repositioning proposition connects with the theory of Pääkkönen et al. (2020), which compares today's algorithmic systems with Michel Crozier's (1964) conceptualization of bureaucratic organizations as complex socio-technical setups where inflexibility is a defining feature. The rigid rules within such setups do not eliminate human discretion but rather redistribute it to areas where uncertainty remains. Similar dynamics occur in algorithmic systems for governance and public administration, where the operation of algorithms cannot yield actionable outcomes without subsequent human judgment. As AI becomes more integrated, human discretion shall remain crucial in managing uncertainties and ensuring responsible decision-making (Alkhatib & Bernstein, 2019; Wang et al., 2022).

Empirical studies also suggest that discretionary practices in social work remain a cooperative endeavorbuilt upon interactions between citizens and bureaucrats rather than arbitrary choice (Flügge et al., 2021; Petersen et al., 2020). Table 1 summarizes selected cases where AI or algorithmic automation tools were integrated into bureaucratic routines in a way that supported, rather than supplanted, human discretion. These examples underscore the importance of institutional norms and human interpretation in shaping discretionary decision-making.

**Table 1**
Summary of selected examples of AI's or algorithmic automation tools' impacts on discretion.

| Case / Study | AI or Automation Application | How Human Agents Use AI in Practice | Mechanisms of Discretional Change | Illustrative Impacts |
|---|---|---|---|---|
| **SALER: Algorithmic Support with Preserved Bureaucratic Discretion** (Criado et al., 2020) | Implementation of SALER, an algorithm-based early warning system to detect financial irregularities in public procurement processes within the Generalitat Valenciana (GVA), Spain. | Civil servants used SALER alerts as starting points but retain full discretion to interpret or override them. They also provide feedback to IT staff to refine risk indicators. | Redistribution of discretion through automation of routine detection and re-emergence of professional judgment at oversight level. | By handling routine detection, SALER preserves human judgment for complex or context-sensitive cases, ensuring decisions remain grounded in professional expertise and situational awareness. |
| **Danish AI-supported Job Placement Services** (Flügge et al., 2021) | AI components are embedded in caseworker tools to predict long-term unemployment risk, providing suggestions on program placements and interventions. | Caseworkers consult risk scores during counselling sessions, using them to prioritize clients or justify service plans to supervisors. They also collectively discuss borderline cases to calibrate interpretation of algorithmic outputs. | Redistribution of discretion through integration of algorithmic recommendations into collaborative routines. Caseworkers translate predictions into professional reasoning rather than following them mechanically. | Caseworkers retained discretion in accepting or rejecting AI suggestions. AI helped save human discretion for more complex, uncertain, or ethically sensitive cases by handling simple tasks or offering information for caseworkers' conversations with supervisors or clients. |
| **Swedish Automated Social Assistance Management** (Ranerup & Svensson, 2023) | Robotic Process Automation (RPA) and algorithmic systems are integrated into the case management routines for social benefit applications in two Swedish municipalities. These technologies automate routine, repetitive tasks by generating suggested decisions based on structured data. | Caseworkers supervise automated workflows, validate system outputs, and intervene when cases require qualitative assessment or exceptions. They adjust rule sets in collaboration with IT staff when routines misclassify applications. | Redistribution of discretion through division of labor between automated verification and human supervision. Judgment persists at points of exception and oversight. | Caseworkers exercised "digital discretion" by navigating between automated routines and human judgment. Caseworkers retained final decision-making authority in situations requiring professional judgment, such as inconsistent application data, client-specific contexts, or discussions around self-sufficiency goals. |

While the focus of this study is on AI-assisted decision-making, it is important to note that some of the algorithmic automation tools used in public administration are often deployed alongside or embedded within AI systems. Although rule-based automation do not possess learning capabilities, its practical integration with AI in hybrid systems can still shape the exercise of discretion. Acknowledging the blurred boundary between traditional ICT, automation tools, and contemporary AI in public sector use cases (van Noordt & Misuraca, 2022), we include selected examples that reflect this convergence. Our emphasis remains on how these systems, be it predictive, supportive, or procedural, impact bureaucratic decision-making in practice.

Across these administrative settings, algorithmic tools formalize repetitive tasks yet reintroduce human judgment at interpretive, supervisory, and exceptional stages. Frontline workers, such as inspectors, caseworkers, and social-assistance officers, all continue to exercise judgment, though increasingly through interaction with digital interfaces and datasets. This continuity suggests that AI alters the organization of discretion more than its existence, embedding human agency within structured, data-driven routines. Such redistribution of discretion, while preserving a human role, also produces new dependencies on technical systems and managerial rules.

However, when embedding AI or algorithms into governmental systems, such technologies can introduce new risks. First, AI systems can disrupt incentive structures. For instance, caseworkers noted that predictive scores encouraged attention to "easier" clients, thereby undermining equity goals by incentivizing work efficiency over needs (Flügge et al., 2021); this finding also echoes with evidence from Swiss public employment agencies (Körtner & Bonoli, 2021). Second, algorithms can create confusion or reduce trust when their inner workings are opaque. Spanish inspectors using the SALER early-warning tool reported difficulty interpreting model logic and uncertainty over responsibility (Criado et al., 2020). The lack of interpretability can undermine trust, prompting resistance or selective disregard of AI-generated analysis or information (Pruss, 2023). Third, discretion may be displaced into less visible spaces, thus reducing accountability. Brayne and Christin (2021)) show how legal professionals resisted algorithmic oversight through subtle practices like data obfuscation and foot-dragging, seeking to preserve autonomy in the face of managerial control. Lastly, over-reliance on algorithmic recommendations risks devaluing experience-based judgment and narrowing reflective practice. As discretion becomes embedded in digital routines, frontline staff may focus on procedural compliance rather than contextual fairness or empathy (Ranerup & Svensson, 2023). These examples illustrate that when algorithmic tools are integrated into existing institutional structures, they can result in practices that privilege procedural efficiency or symbolic compliance over public values. To address these risks, AI must be governed through robust oversight and sustained investments in capacity-building to ensure it augments rather than overriding professional judgment (as discussed in Section 4).

### 3.3. *Reshaping accountability in AI-assisted urban governance*

In public administration, accountability is essential for maintaining public trust and ensuring effective governance. Data-driven and AI-assisted approaches introduce new concerns to accountability, particularly in ensuring explanation, justification, fairness, and oversight for AI-driven decisions (Busuioc, 2021; Novelli et al., 2023). These concerns can be viewed from two key angles: *algorithmic accountability*, referring to that AI systems themselves operate transparently and ethically, and *accountable AI use*, referring to how human agents, assisted by AI, maintain accountability in their decision-making processes. For the former, the inherent opacity of algorithmic processes makes it difficult to understand and justify AI-driven decisions (Bullock, 2019; Zuiderwijk et al., 2021; Busuioc, 2021). For the latter, because interpreting algorithmic outcomes remains technically difficult, scholars have turned to governance approaches as a way to secure explanations and justifications for AI-assisted decisions (Liu et al., 2019; Zuiderwijk et al., 2021; Dwivedi et al., 2021). These governance approaches typically involve implementing external oversight, designing clear explanation criteria, establishing procedures for hearings, and enforcing consequences for non-compliance (Busuioc, 2021).

To further explore how AI reshapes accountability within bureaucratic systems, it is useful to distinguish among three types of public accountability: *political, professional,* and *participatory* (Hupe & Hill, 2007). *Political accountability* characterizes vertical relationships between street-level bureaucrats and their supervisors; *professional accountability* focuses on horizontal accountability to peers and professional associations, adhering to professional standards and norms; and *participatory accountability* centers on client-based evaluations and feedback from citizens as well as collaborative decision-making processes among citizens and public officials. Each form of accountability may necessitate a different mechanism for AI-driven oversight and control, allowing for a more nuanced approach that respects the need for both discretion and accountability for context-specific challenges.

### 3.3.1. Political accountability

Deploying highly interactive AI applications has the potential to reinforce top-down authority by enhancing managerial control over frontline behaviors. In this framework, AI may facilitate data-driven oversight and evaluations of bureaucratic decisions, thereby enhancing political accountability. Officials have traditionally used data to understand and monitor patterns of rules, decisions, and behaviors to safeguard accountability and performance (Janssen et al., 2020). Algorithmic systems extend this logic by generating continuous traces of decisions that make frontline work more visible and auditable.

A compelling illustration comes from the Brazilian Environmental Military Police, where officers transitioned from face-to-face “street” policing to “screen” policing, supported by data-driven analytics and GPS tools to assist interventions (Aviram et al., 2024). These tools integrate data from field operations, enabling commanders to observe patrol movements, review intervention logs, and evaluate adherence to protocols in real time. At the same time, street officers reported that algorithmic dashboards changed how they prioritized patrols by emphasizing measurable indicators such as patrol frequency or proximity to incidents over contextual judgment. Such datafication of performance illustrates how political accountability can expand managerial control while subtly reshaping professional discretion. Comparable dynamics appear in other administrative settings as discussed previously. Spain’s SALER early-warning system also produces auditable, explainable outputs aligned with inspectors’ work questions, which senior officials can use to flag anomalies and justify oversight actions, thereby enhancing upward visibility (Criado et al., 2020).

The recent emergence of GenAI tools may accelerate the transition from traditional “street” practices to data-driven “screen” practices. Most applications based on Large-Language Models (LLMs) can respond to plain language requests, significantly lowering the hurdle for frontline caseworkers and their supervisors to analyze data in their decisions. AI has also demonstrated strong capabilities to organize unstructured inputs into meaningful insights (Williams et al., 2024; Gupta et al., 2025) as well as to detect abnormal patterns in behaviors or decision-making (Jassen et al., 2020; Criado et al., 2020). By producing clear, auditable trails of decisions and actions, these tools demonstrate potentials to enhance transparency and reduce concerns about arbitrariness or bias, but they also extend managerial oversight deeper into day-to-day operations.

However, heightened monitoring introduces costs. While it enables greater top-down accountability and protects against misconduct, it can also reduce officers’ trust in technology, especially if tools are perceived as instruments of control rather than support (Kellogg et al., 2020; Aviram et al., 2024). The same visibility that strengthens accountability can narrow professional judgment. Effective political accountability, therefore, must balance oversight with professional autonomy to sustain discretion and flexibility in frontline work. As AI continues to reshape bureaucratic visibility, new forms of work relationships and positions will emerge, while carrying both the legacy of hierarchical control and the possibilities of data-enabled collaboration (Giest & Klievink, 2024; Kellogg et al., 2020). This dynamic shift sets the stage for Section 4.2, which advocates adaptive administrative structures for AI adoption.

### 3.3.2. Professional accountability

Professional accountability refers to adherence to professional standards and norms, ensuring that bureaucrats perform their duties competently and ethically, and in alignment with public values (Hupe & Hill, 2007; Moore, 2013). Professional expertise may serve as a safeguard against both arbitrary rule-following and personal bias. As Evans (2011) and Maynard-Moody and Musheno (2000) note, street-level bureaucrats often view themselves as autonomous decision-makers whose discretion depends on deep expertise and ethical considerations. In this context, AI can assist street-level bureaucrats by standardizing critical processes, improving procedural fairness, and aligning goals while still allowing for professional discretion where necessary.

AI-driven recommendations, either requested by human agents or integrated into on-site applications, can generate prompt, scenario-specific suggestions, reducing variability and personal biases (Selten et al., 2023). In fields like social work, predictive or assistive tools can help align day-to-day practice with organizational goals, fostering procedural justice and consistency (Bullock, 2019; Bullock et al., 2020; Young et al., 2019). When designed collaboratively, such tools can also support continuous learning and professional development (Cukurova et al., 2024).

The case of two Swedish municipalities, previously mentioned, offers a clear example of how AI-enabled systems can reinforce professional accountability (Ranerup & Svensson, 2023). In Mölndal, Robotic Process Automation (RPA) handled routine eligibility assessments, allowing caseworkers to concentrate on client engagement. Standardized e-applications and automated workflows also enhanced the consistency and reduced errors in decisions. In Malmö, these tools created capacity for more frequent follow-up meetings, supporting a shift toward personalized, proactive care. Crucially, local developers and frontline caseworkers collaborated on the design and calibration of automated rules. That gave caseworkers a sense of ownership and reinforcing their role as custodians of ethical and professional quality. The redistribution of responsibility from repetitive work such as tedious data entry toward oversight, interpretation, and client care demonstrates how AI can strengthen, rather than weaken, professional accountability.

However, risks persist. First, when automation is perceived as opaque or rigid, professional public workers may resist any technological tool perceived as undermining professional expertise or limiting case-sensitive flexibility (Kellogg et al., 2020; Pruss, 2023; Stevenson & Doleac, 2024). Second, managerial pressures for efficiency can conflict with practitioners’ commitment to client-centered service (Evans, 2011; Wang et al., 2022). Third, unequal participation in system design may erode the sense of ownership that underpins accountability, reducing AI from a collaborative partner to an imposed control mechanism. These challenges underscore the need for adaptive administrative structures (as discussed in Section 4.2) and transparent human–AI collaboration (as discussed in Section 4.4).

### 3.3.3. Participatory accountability

While frontline bureaucrats mostly either facilitate citizens’ interactions with governmental agencies (e.g., obtaining a license, paying a fee) or regulate citizens’ actions (e.g., stopping speeding), with established feedback and client-based evaluations, frontline bureaucrats can also be held accountable by citizens (Hupe & Hill, 2007). Beyond being subjects of regulation or service delivery, citizens also act as accountability agents by questioning, contesting, and co-producing public services. As interactive AI technologies diffuse, they may increasingly mediate this two-way accountability, allowing both bureaucrats and citizens to access and interpret information that once flowed in only one direction.

Early pilots include chatbots and virtual assistants that translate complex administrative language into conversational formats and provide round-the-clock access to public information. The review by van Noordt and Misuraca (2022) finds that chatbots and virtual assistants are the most common form of AI system across European public administrations (57 of 250 cases), with >90 percent deployed for service-delivery purposes. Larsen and Følstad (2024) illustrate how the Norwegian municipal chatbot “Kommune-Kari” provides citizens with 24/7 self-service access to local government information and services. Municipal staff then analyze anonymized chatbot logs to better understand recurring pain points and refine processes. This dual function: citizen guidance and institutional learning, creates a continuous feedback loop that strengthens transparency and responsiveness. Similarly, through over 507 sets of responses, Pislaru et al. (2024) show how AI-based virtual assistants significantly increase citizens’ willingness to interact with public agencies, especially among digitally literate groups. By simplifying access to policy information, virtual assistants facilitate more frequent and informed inquiries, and their back-end analytics enable officials to detect recurring bottlenecks in service delivery. These examples show how conversational AI tools can transform citizen–government communication into an iterative exchange rather than a one-way request–response process.

Such enhancements reflect a form of participatory accountability. Empowered with AI-enabled communication tools, citizens not only receive more respectful and comprehensible replies but also gain improved capacity to interpret government messaging critically and provide informed feedback. In essence, the AI-enabled interaction structure allowed citizens to better scrutinize government actions while simultaneously being reassured by clearer, more empathetic responses. Furthermore, community-based organizations (CBOs), in particular, can harness such tools, along with government’s open data and policy information, to identify patterns, make cross-neighborhood comparisons, and hold local agencies accountable. Thus, AI may not only improve the quality of bureaucratic responses but also enable both citizens to participate more meaningfully in oversight and governance.

The potential of AI to democratize access to information and enhance public engagement is substantial, but it also leads to challenges for both citizens and local governments. First, for citizens, the ease of access to large volumes of policy information risks overwhelming individuals with too much convoluted information, leading to confusion rather than clarity. Second, education-linked disparities in AI literacy and digital skills could exacerbate existing inequalities in who gets to participate effectively (Pislaru et al., 2024), leading to situations where those with more resources and knowledge are empowered to interact more effectively with government, while others being further marginalized. Third, privacy concerns may deter citizens from engaging, especially when it is unclear how personal data collected through AI platforms is used or protected.

On the government side, a significant challenge lies in the varied data-driven expertise among public officials. The skill gap is twofold: there is a disparity between individuals with little to no data knowledge and those with some level of data expertise, and even among those with data skills, proficiency can vary significantly. This uneven landscape hampers effective collaboration and impedes the ability to critically assess and integrate AI-generated insights with professional judgment, as observed in Section 3.2, where AI implementation demanded new interpretive skills from caseworkers and supervisors (see Flügge et al., 2021; Ranerup & Svensson, 2023). Second, rising expectations for responsiveness and personalization can easily outpace the actual capacity of public agencies to deliver, thereby potentially undermining public trust. Lastly, many local governments may struggle to process and respond to the influx of citizen inputs generated through AI-enabled systems. These tensions remind us of that participatory accountability depends as much on institutional inclusion and capacity as on technological design. Equitable deployment, transparent data governance, and human-centered implementation are therefore essential to ensure that AI-enabled citizen engagement enhances, rather than fragments, public oversight and trust (as discussed in Section 4).

### *3.4. Accountable discretion: balancing discretionary flexibility with oversight*

AI can play a significant role in reshaping the dynamics between discretion and accountability by enhancing the government’s internal technical capacity (see Fig. 1). However, enhanced capacities across different bureaucratic levels, in turn, yield different impacts. For frontline bureaucrats, as discussed in previous sections, these enhanced capacities are likely to manifest through offloading routine tasks, aligning operational objectives, and ensuring adherence to broader regulatory and ethical frameworks. For example, a social worker handling a case involving a family with multiple socio-economic issues can use AI to analyze historical data, available assistance, and previous case notes before finalizing potential interventions and resources. This collaborative approach ensures that experienced case workers retain their discretion in decision-making while also making sure that decisions meet operational objectives and adhere to organizational values and guidelines.

For managers, AI can significantly enhance oversight capabilities by processing extensive amounts of data in either real-time or ex-post reviews. As suggested in the previous sections, rather than reiterating traditional accountability challenges, managers now have the capacities to compare decisions against established protocols, legal standards, and ethical norms through automated detection systems that flag deviations promptly. This immediate feedback loop enables swift intervention with corrective actions when necessary.

In sum, as detailed above, AI enables a shift where routine tasks are automated and human judgment is reserved for complex decision-making. This shifted balance is maintained through discretionary flexibility and robust oversight measures, ensuring decisions remain both professional and accountable.

### *3.5. Acknowledging the technical challenges of embedding ethics into AI*

Accountable discretion presupposes specific technical and institutional conditions. AI’s potential contributions to reshaping discretionary decision-making and ensuring accountability in governance come with significant technical and normative challenges. The first challenge is to ensure that AI systems are traceable and capable of providing justifications for their decisions (Busuioc, 2021; Novelli et al., 2023). AI systems often operate as "black boxes," making it difficult for users to understand how decisions are made (Adadi & Berrada, 2018; Došilović et al., 2018; OECD 2024; Chen et al., 2025). When explanations are inaccessible, human officials lose the ability to question, override, or contextualize automated outputs. Addressing this challenge requires institutional measures that keep human agents as proactive decision-makers and embed interpretability and contestability into AI deployment, ensuring that models remain intelligible to auditors (see Section 4.4).

Another major challenge is how to operationalize public values into rules that can be programmed and applied consistently across foundational models and their downstream applications (see Dwivedi et al., 2021; Sun & Medaglia, 2019; Wirtz et al., 2019; Bommasani et al., 2023). Algorithmic applications in high-stakes domains must be assessed not only for fairness and transparency but also for their broader societal implications (Mittelstadt et al., 2016). Yet, attempts to operationalize ethics through design parameters can narrow normative reasoning into engineering procedures. For instance, welfare or risk-assessment algorithms, if automated, risk compressing fairness principles into numeric thresholds. That may turn judgments into optimization problems. When such moral reasoning is recast as logic that is encoded in modeling, metrics, or procurement requirements, questions of accountability migrate from public agencies' deliberation to technical design, reshaping who exercises authority.

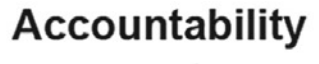

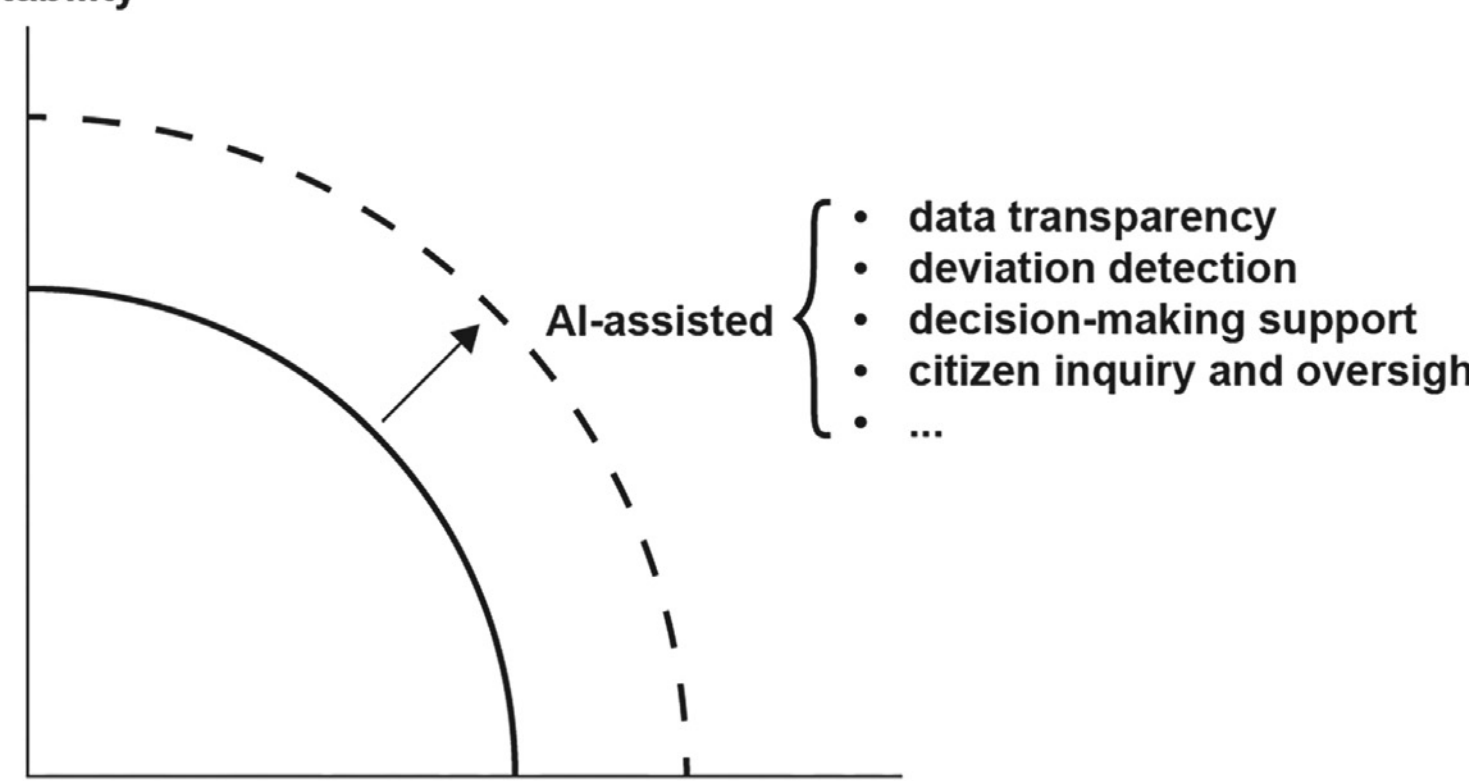

**Fig. 1.** . With enhanced data transparency, real-time monitoring, and algorithmic decision support, the introduction of AI has the potential to shift the trade-off boundary outward, thereby enhancing both accountability and discretion. Source: Authors' own diagram.

In public administration, this dynamic can shift decision-making away from governments toward private-sector model developers and vendors, making accountability more complex rather than more transparent. These tensions resonate with the broader "techlash" discourse, which, while initially expressing public anxiety against large technology corporations, has evolved into a sustained critique of private power and surveillance capitalism (Helles & Lomborg, 2024).

These concerns are compounded by the fragmented and opaque AI supply chain through which most public-sector applications are developed and deployed. This supply chain spans private vendors, public-sector IT units, and open-source contributors (Cen et al., 2023; Cobbe et al., 2023). In this context, the concentration of innovation power among a few leading technology corporations further complicates governance and intensifies risks to regulation and oversight (Hacker et al., 2023; Widder & Nafus, 2023; Narechania & Sitaraman, 2023). At the city-level, when governments treat AI adoption as a technical procurement rather than a policymaking endeavor, they risk ceding value-laden decisions to algorithm designers beyond public scrutiny (Mulligan & Bamberger, 2019).

That said, embedding ethics in AI therefore demands more than technical standards. It requires institutional safeguards that maintain public accountability throughout the entire AI development and deployment pipeline. The following Section 4 elaborates on the institutional responses to these challenges ranging from administrative capacity building and adaptive governance structures to citizen-engaged oversight mechanisms.

## 4. Guiding principles for AI-enhanced governance in urban policy

While Section 3 discusses how AI can reconfigure discretion and accountability, this section turns to guiding principles for governing these transformations. Rather than prescribing a one-size-fits-all blueprint, this section proposes a set of guiding principles as a heuristic framework that urban jurisdictions can calibrate to their own local regulation, capacity, labor relations, and politics in order to cultivate *accountable discretion* – that is, the ability to exercise judgment flexibly while maintaining transparency and ethical control in AI-assisted governance practices.

Each principle captures a key dimension of governing with AI: promoting equity in capability building, enabling adaptive administrative structures, ensuring robust data governance, embedding proactive human oversight, and engaging citizens in continuous review. Table 2 summarizes the common barriers and facilitators that influence whether these principles can be effectively translated into day-to-day practice across diverse administrative settings. The following Sections (4.1–4.5) examine each principle and discuss how governments can adopt them to cultivate accountable discretion in their specific urban settings.

### *4.1. Equitable AI deployment: ensuring uniform accountability in urban governance*

Principle 1: *AI should be deployed in a manner that ensures equitable enhancement of capabilities across government roles, so that all employees, irrespective of their technical background, can benefit uniformly from AI integration, thereby reinforcing accountability.*

Disparities in AI literacy and access, among both bureaucrats and citizens, risk entrenching existing inequalities (Agostino et al., 2022; Goelzer, 2023). For government operations, AI applications do more than simply modularize discretionary decision-making at the street level; they amplify the capacities of individual governmental employees. Yet, AI's magnifying effect is likely to be distributed unevenly. Those with stronger backgrounds in data analysis and technical skills are better positioned to benefit, while employees without formal training may face steeper barriers to adoption. Such internal disparities can shift decision-making power across hierarchical lines by possibly favoring managers and central offices and risking disruptions to the existing accountability structures. These concerns become even more pressing when AI is applied in citizen-facing contexts. Frontline caseworkers serving marginalized communities may themselves lack adequate access to AI tools or training, while working with residents who face digital exclusion and low AI literacy. While some CBOs and advocacy groups may use AI applications to interpret open data and represent community interests, widespread disparities in AI and data literacy persist, particularly among those not connected to such organizations. The current ability to meaningfully engage with open data or algorithmic systems remains uneven, and without targeted support, there is a risk that AI may widen gaps in service access and participatory oversight.

To mitigate these risks, public agencies should institutionalize equity-oriented capacity building. First, governments should assess existing skill distributions through organization-wide audits and identify priority areas for training. Second, they should organize initiatives, such as workshops, mini-lectures, and lunch talks, that enhance AI literacy, data interpretation, and AI risk awareness (Allen et al., 2024). Third, agencies could embed role-specific tutoring courses that reflect the distinct responsibilities of managers, analysts, and frontline staff, combining peer-support networks and mentoring programs that promote knowledge sharing across departments, ensuring that expertise does not concentrate within a few technically fluent units. Finally, governments should ensure equitable access to digital infrastructure and analytical tools, preventing resource disparities between central offices and frontline operations.

**Table 2**
Barriers and facilitators for implementing AI governance principles.

| Guiding Principle | Common Barriers | Facilitators for Implementation |
|---|---|---|
| *4.1 Equitable AI Deployment* | • Uneven AI and data literacy among general and domain-specific staff<br>• Limited time and incentives for training<br>• Unequal access to digital tools and infrastructure | • Conduct organization-wide audits to map skill and resource gaps<br>• Provide role-specific AI-literacy programs for different staff roles<br>• Establish peer-support or mentoring ("AI champions")<br>• Ensure equal access to digital tools and analytics platforms |
| *4.2 Adaptive Administrative Frameworks* | • Rigid job classifications and outdated civil-service rules<br>• Union or labor-agreement constraints on role flexibility<br>• Departmental silos and fragmented data ownership<br>• Risk-averse culture and/or limited incentives to experiment | • Redesign job roles to include AI-related responsibilities<br>• Establish cross-agency AI task forces or project "squads"<br>• Create joint labor-management pilot programs<br>• Introduce safe, time-limited innovation sandboxes<br>• Incorporate adaptive performance and/or innovation practice indicators that reward learning |
| *4.3 Robust Data Governance* | • Privacy, consent, and legal uncertainties<br>• Incomplete or poor-quality data inventories due to legacy IT systems and fragmented data standards<br>• Limited auditing and stewardship capacity for oversight | • Establish data-protection principles aligned with in-practice legal frameworks, such as OECD AI Principles, GDPR, EU AI Act, etc.<br>• Adopt privacy-by-design measures (encryption, minimization, access control)<br>• Conduct comprehensive data inventories and audits<br>• Adopt shared metadata conventions / open APIs for interoperability<br>• Create data-stewardship teams and algorithm registers |
| *4.4 Proactive Human Judgment and Institutional Mechanisms* | • Over-reliance on automated outputs ("automation bias")<br>• Symbolic documentation, inconsistent audits, and weak post-deployment monitoring<br>• Lack of clear override or escalation procedures | • Establish criteria distinguishing human vs. automated decision domains<br>• Use documentation tools (model cards, datasheets) and maintain algorithm registers and conduct impact assessment on AI deployment<br>• Create internal ethics boards and escalation channels for override or appeal<br>• Conduct periodic reviews for model drift and ethical impact<br>• Provide training in ethical reasoning, bias awareness, and deliberative judgment |
| *4.5 Engaging Citizens in Oversight* | • Low public AI literacy and limited awareness of algorithmic systems<br>• Digital divide and unequal access<br>• Information overload or consultation fatigue | • Publish non-technical AI summaries through public dashboards or registers<br>• Develop citizen-facing tools (e.g., chatbots, explainers) that both inform and collect feedback<br>• Partner with community organizations to reach under-represented groups<br>• Provide accessible, multilingual interfaces and report back on how citizen input is used |

A shift toward tech-based capacity building and inclusion is crucial for reinforcing managerial control and public oversight (Engstrom & Haim, 2023). By ensuring that all employees can engage meaningfully with AI tools, governments can create the human-centric foundation that allows adaptive administrative structures to emerge (see Section 4.2); in turn, those adaptive structures are what sustain equitable AI deployment in the long run. Such reciprocal dynamics help anchor accountable discretion within AI-enhanced urban governance.

### 4.2. Adaptive administrative frameworks for AI integration

Principle 2: *Civil service frameworks and union agreements must be reformed to enable innovative and adaptive administrative structures that can effectively integrate AI technologies, ensuring that public organizations remain agile and responsive in a rapidly evolving technological landscape.*

Traditional civil service job classifications and strict union agreements have long provided stability and fairness in public employment. However, as AI and other digital technologies reshape urban governance, these rigid structures can hinder the necessary flexibility and cross-functional collaboration. Many job descriptions still exclude responsibilities for data oversight or model evaluation, and collective-bargaining rules can restrict the reassignment of duties linked to AI implementation. In unionized labor contexts, collective bargaining work rules can restrict the redefinition of job roles and the incorporation of new tasks related to AI, slowing down the implementation process. Such rigidity limits the ability of governments to experiment, iterate, and respond quickly to technological change.

A practical approach to building adaptiveness can unfold in several stages. First, governments should revise job descriptions and staffing frameworks to include emerging responsibilities such as AI management and data stewardship, bridging the gap between existing civil-service competencies and new technological needs. Second, they can establish cross-departmental AI task forces or "innovation squads" empowered to test digital tools in controlled settings and share lessons learned, as demonstrated by the Boston Department of Innovation and Technology's recent experiments (Williams et al., 2024). In parallel, pilot agencies (ideally supported by executive leadership) can negotiate pilot clauses within labor agreements that permit time-limited experimentation while preserving job security and fairness. Third, once such institutional flexibility is established, other departments can create innovation sandboxes to trial AI applications under shared ethical and performance standards, building trust through small-scale, low-risk experiments. It is equally important that these structural changes are implemented in close dialogue with union representatives, ensuring that the necessary flexibility does not undermine job security or the foundational fairness of public employment (Engstrom et al., 2020).

Sustained adaptiveness also depends on a culture of continuous learning that rewards experimentation and reduces fear of failure. Integrating AI-related competencies into professional-development programs and recognizing innovation in promotion criteria can help institutionalize adaptability as part of bureaucratic professionalism. Over time, adaptive structures and equitable capacity-building (see Section 4.1) may reinforce one another in a way that a skilled and confident workforce can operate effectively within flexible frameworks, and such frameworks, in turn, enable ongoing skill renewal. These adaptive mechanisms ultimately depend on the quality and accessibility of underlying data systems, which form the backbone of accountable AI governance, as discussed in Section 4.3.

### 4.3. Robust data governance: the backbone of accountability in AI-assisted urban governance

Principle 3: *Reliable and ethically governed data is essential for ensuring accountability in AI-enhanced urban governance. Robust data management practices must balance the need for institutional control with the coordination required to make AI systems transparent, fair, and auditable.*

Two foundational priorities, protecting data privacy and ensuring data integrity across institutions, capture the minimum conditions for the balance between professional judgment and institutional oversight. The first priority concerns ethical safeguards and privacy protection. AI applications in public services often require extensive data collection. These issues are especially acute in contexts where digital systems expand under the justification of efficiency, potentially leading to "surveillance creep" that compromises individual autonomy and trust (Eubanks, 2018). To address these risks, first, local governments should establish clear data protection and governance principles that define what personal data may be collected, under what conditions they can be used, and how long they may be retained. These principles should be aligned with foundational legal framework (e.g., OECD AI Principles, GDPR, EU AI Act, etc.) and ensure that individuals are informed about the use of their data in training and fine-tuning AI systems. Next, domain-specific agencies should embed privacy-and security-by-design standards into both procurement and development processes for AI systems. That means incorporating measures such as data minimization, anonymization, and access control from the outset rather than adding them retrospectively. Finally, local governments should develop transparent communication and contestation mechanisms that explain how AI tools use data and enable government employees and citizens to exercise their rights to access, correction, or erasure under data regulation. While these measures are not unfamiliar in data governance practices, they do more than ensure regulatory compliance. They make decision-making legitimate and accountable in the eyes of the public by ensuring that the boundaries of data use are visible and that both human and algorithmic decision-making remain explainable.

The next priority concerns data quality and interoperability. Fragmented data infrastructures and poor data quality can result in significant adverse outcomes, including discrimination, political polarization, and the erosion of public trust in AI-driven processes (Angwin et al., 2016; Simons, 2023). To address these challenges, agencies should begin by conducting comprehensive audits to assess the completeness and accuracy of existing datasets. Next, governments should adopt shared metadata standards and open APIs that enable interoperability, secure data exchange, and cross-departmental benchmarking. In parallel, these interoperability mechanisms should also incorporate monitoring routines (e.g., periodic quality reviews, documentation of data provenance, and clear procedures for error correction) to ensure data integrity. Finally, Complementary training in data stewardship and statistical literacy helps staff link these technical practices to ethical implications (Agostino et al., 2022; Janssen et al., 2020). These actions ensure that data remain interpretable and contestable rather than opaque or siloed.

It is critical to note that these measures are interconnected rather than sequential. Technical reliability, legal compliance, and public accountability reinforce one another, and their coordination depends on robust data-governance frameworks that link technical, legal, and managerial capacities across departments. In many administrations, this coordination is supported by dedicated data-stewardship teams or equivalent offices, whose institutional form varies by context (e.g., San José's Digital Privacy Office, Barcelona's Municipal Data Office, Boston's Department of Innovation and Technology). Such units could help ensure that audits, validation routines, and ethical reviews are consistently implemented. Ultimately, effective data governance converts reviewable data systems into an evidentiary foundation for the proactive human judgment and oversight mechanisms discussed next.

### 4.4. Proactive human judgment and institutional mechanisms for accountability

Principle 4: *Effective human–AI collaboration requires proactive accountability mechanisms that center human oversight in AI-assisted decision-making. While AI can support data processing and analysis, human judgment must remain the final authority to mitigate algorithmic opacity, uphold ethical standards, and prevent over-reliance on automation.*

The evolving integration of AI into urban governance is transforming the relationship between human agents and automated systems into a dynamic, colleague-like partnership (Meijer et al., 2021; Engstrom & Haim, 2023). In this kind of partnership, ensuring that human discretion remains central in AI-assisted governance involves three mutually reinforcing strategies. First, public authorities must have clearly defined guidelines that establish the boundaries of AI's role. These guidelines should consider the stakes, reversibility, and ethical implications of each decision type. Clear documentation of "human-in-the-loop" and "human-on-the-loop" roles helps delineate accountability and ensures that oversight is both traceable and defensible (Engstrom & Haim, 2023). For instance, as observed in Sweden's automated social-assistance management, routine eligibility checks were automated, but final decisions and client interactions remained human-led (Ranerup & Svensson, 2023).

The second strategy is to institutionalize regular review and evaluation of AI systems. Effective human oversight depends on procedural mechanisms that allow systems to be questioned and corrected over time. Some notable examples include Algorithmic Impact Assessments (AIAs), standardized documentation tools, such as model cards and datasheets, and public algorithm registers (as created in Helsinki and Amsterdam, see Floridi, 2020), all of which exemplify institutional responses that aim to improve AI interpretability, data transparency, and expose potential risks (Reisman et al., 2018). However, these tools may still risk becoming symbolic practices if exercised without regulatory support or organizational commitment. For instance, AIAs often rely on voluntary cooperation from firms, which may be undercut by liability concerns and profit motives (Selbst, 2021); standardized algorithmic documentation could foster "false assurance" if audits lack independence or rigorous methodology (Goodman & Trehu, 2023); public algorithm registers are typically voluntary, often omit high-stakes use cases (like policing or fraud detection), and rarely incorporate concrete channels for citizen input or contestation (Cath & Jansen, 2021; Goelzer, 2023). Therefore, oversight should not be treated as a compli-ance formality but as a continuous accountability practice supported by periodic reviews, multi-stakeholder evaluation, and escalation channels.

Third, governments should build effective feedback and escalation mechanisms that empower human decision-makers. Public officials using AI tools should have clear channels to question, override, or flag questionable outputs without penalty. That should include establishing review committees or ethics boards to handle contested cases, as well as technical affordances that allow users to document their reasoning when diverging from algorithmic recommendations. In criminal justice settings, for example, professionals have adopted subtle forms of resistance, such as delaying or selectively disregarding algorithmic advice, or recording alternative rationales, to preserve interpretive autonomy when system outputs conflicted with their experience-based judgment (Brayne & Christin, 2021). Such feedback and resistance practices help maintain professional discretion while promoting reflective organizational learning.

These strategies aim to ensure that AI systems support rather than supplant human agency. However, these mechanisms are only effective when embedded within an organizational culture that values professional judgment and deliberation. Professional development and AI-literacy training that emphasize ethical reasoning, fairness, and bias awareness can help cultivate such a culture. Over time, proactive human oversight and the preceding principles (see Sections 4.1–4.3) could reinforce one another, with equitable capacity-building supporting responsible decision-making and robust data governance providing the evidentiary foundation for oversight. Beyond these institutional conditions, meaningful citizen engagement, as discussed next in Section 4.5, also remains essential to ensuring AI systems serve the public interest.

### 4.5. Citizen oversight in AI-assisted urban governance

Principle 5: *City governments should leverage their close connection to local communities to strengthen participatory accountability by building a reciprocal mechanism of oversight, in which citizens not only understand how algorithmic systems affect them but also influence their design, monitoring, and improvement. Engaging residents in these processes fosters transparency, mutual trust, and alignment between AI applications and public values.*

City-level governments occupy a unique position in urban governance, given their continuous and direct engagement with local communities. This proximity advantage provides a stronger public preference for local over national AI applications, which can be harnessed to build trust (Schiff et al., 2023). By capitalizing on this trust, local governments can translate technical systems into accessible terms and address concerns about data privacy, bias, or automation (Lee, 2018; Araujo et al., 2020), while also incorporating citizen feedback into the deployment and evaluation of AI tools. As discussed in Section 3.3, participatory accountability is a two-way relationship of explanation and response. In this sense, local engagement complements governmental accountability by ensuring that discretion exercised by frontline bureaucrats remains aligned with public values.

Participatory accountability in the AI context can be advanced through two mutually reinforcing pathways. The first is to improve accessibility and transparency of information. Local governments should publish nontechnical summaries of AI projects via dashboards, open-data portals, or algorithmic registers to communicate what systems are in use, their purposes, and their limitations. Government chatbots illustrate how conversational interfaces help citizens navigate public services and comprehend administrative processes, while also enabling agencies to detect informational gaps and improve communication (Chen et al., 2023; Larsen & Følstad, 2024; Pislaru et al., 2024). Experiments in Boston, where residents used generative-AI tools to analyze planning documents, demonstrate how such transparency mechanisms can deepen public understanding and enable citizens to interrogate AI-generated insights (Williams et al., 2024).

Second, local agencies need to promote inclusiveness. Participatory accountability must ensure that opportunities to scrutinize or contribute to AI systems are open to diverse social groups, not limited to those with technical expertise or privileged access. Governments should identify and mitigate barriers such as low digital literacy, limited connectivity, or language obstacles. Partnerships with CBOs can extend outreach to marginalized populations, while public training and educational programs can help citizens interpret algorithmic explanations and data visualizations.

While earlier sections discussed the practical limits of participatory accountability, ranging from information overload to uneven digital literacy (see Section 3.3), governments can address these challenges through procedural clarity and communication. What matters is not the volume of participation but the credibility of responses. Establishing clear engagement protocols that specify how citizen input informs AI policy or service design, and publicly reporting on resulting actions, can transform participation from one-time consultation into an iterative process of building accountability.

Participatory accountability complements rather than replaces institutional oversight. While the previous principles support discretionary decision-making and secure transparency within government, citizen-engaged oversight ensures that AI systems remain grounded in public legitimacy. Together, these mechanisms could connect bureaucratic discretion, data governance, and citizen voice in sustaining ethical and trustworthy AI-assisted urban governance.

## 5. Conclusion

This paper has examined how the rise of AI, especially interactive and generative systems, reshapes the relationship between bureaucratic discretion and accountability in urban governance. Building on public administration theory and emerging literature on AI in public sectors, we argue that AI does not simply constrain discretion, but rather redistributes and reconfigures it across institutional levels, professional roles, and citizen interactions. Our conceptual framework clarifies how political, professional, and participatory forms of accountability are each uniquely impacted by AI deployment. These shifts expand the relationship between discretion and accountability and allow, in many cases, both to increase simultaneously. For example, AI can empower frontline officials to exercise more informed and context-sensitive judgment, while also equipping supervisors with tools for monitoring behavior and ensuring compliance. Rather than reinforcing a zero-sum trade-off, AI could introduce a new dynamic in which discretion and accountability can be mutually reinforced, though not without risks. In response, we propose five guiding principles to support responsible and adaptive AI integration in government, emphasizing equitable deployment, adaptive administration structures, robust data governance, human-led judgment, and participatory oversight. Together, these contributions offer a novel lens for understanding AI not merely as a technical tool, but as a governance project that demands new forms of institutional design and ethical responsibility. While the paper emphasizes AI's potential, it also identifies persistent risks across its impacts on discretion and different accountability forms, such as managerial overreach, professional resistance, digital inequity, and implementation gaps.

Moving forward, empirical research is essential to validate and refine the framework and guiding principles proposed in this paper, particularly the concept of accountable discretion. Rather than measuring it as a score or percentage, accountable discretion can be observed through patterns in how discretion and accountability are practiced and negotiated between humans and algorithmic systems. Future research can examine how AI reshapes these relationships along two key dimensions: (1) decision latitude and (2) transparency and justification. First, surveys and interviews can capture how public officials perceive their discretionary autonomy, fairness, and accountability, providing insight into the subjective and interpretive dimensions of discretion (for example, Aoki, 2020; de Boer & Raaphorst, 2023Nagtegaal, 2021; Wang et al., 2022). In the meantime, digitized administrative systems, such as case-management, permitting, or customer-service platforms (e.g., Boston's

BOS:311), generate workflow data that record decision timing, approvals, and overrides. These action traces can complement traditional methods by showing how discretion is exercised in real administrative procedures and how algorithmic recommendations are followed, modified, or contested. However, using such data from digital administrative systems often remain difficult due to restricted access and inconsistent documentation. Research feasibility will depend on data transparency and institutional willingness to share information, and qualitative inquiry will remain indispensable for interpreting administrative action traces and situating them in their institutional and ethical contexts.

To translate these ideas into testable hypotheses, future research should adopt mixed-method designs. Such studies might ask:

1. How does AI deployment for decision-support and performance monitoring alter frontline officials' perceived decision latitude and managers' oversight practices?
2. Under what institutional conditions do AI-assisted oversight practices enhance or constrain professional judgment?
3. How do internal documentation practices and external transparency tools (e.g., algorithm registers, disclosure, decision notice, etc.) reshape the perception of responsibility and legitimacy among bureaucrats and citizens?

Addressing these questions through empirical cases would help test the mechanisms proposed in this paper and clarify whether AI empowers institutional capacity by fostering accountability through more informed and transparent discretion.

## Funding statement

This research received no specific grant from any funding agency in the public, commercial, or not-for-profit sectors.

## Declaration of competing interest

The author(s) declare no conflicts of interest.

During the preparation of this work the author(s) used Grammarly in order to conduct proofreading and readability improvement. After using this tool/service, the author(s) reviewed and edited the content as needed and take(s) full responsibility for the content of the publication.

## CRediT authorship contribution statement

**Stephen Goldsmith:** Writing – review & editing, Supervision, Conceptualization. **Juncheng Yang:** Writing – review & editing, Writing – original draft, Visualization, Methodology, Formal analysis, Conceptualization.

## Acknowledgement

We would like to thank Betsy Gardner and Dr. Quinton Maynefor their advice and suggestions during the revision process.